\documentstyle[twocolumn,prb,aps,epsfig]{revtex}

\begin{document}
\title{
       Anomalous thermal conductivity of frustrated Heisenberg
       spin-chains and ladders}

\author{J.V. Alvarez and Claudius Gros
       } 
\address{Fakult\"at 7, Theoretische Physik,
 University of the Saarland,
66041 Saarbr\"ucken, Germany.}

\maketitle

\begin{abstract}
We study the thermal transport properties
of several quantum spin chains and ladders. 
We find indications for a diverging thermal conductivity at finite
temperatures for the models examined.
The temperature at which the non-diverging prefactor
$\kappa^{(th)}(T)$ peaks is in general substantially lower
than the temperature at which the corresponding 
specific heat $c_V(T)$ is maximal. We show that this result
of the microscopic approach 
leads to a substantial reduction for estimates of
the magnetic mean-free path $\lambda$ extracted by analyzing 
recent experiments, as compared to similar analyses
by phenomenological theories.
\end{abstract}
PACS numbers:


\vspace*{1cm}
{\bf Introduction} -
The nature of thermal transport in systems with
reduced dimensions is a long standing problem and
has been studied intensively in classical~\cite{bon00} 
systems either by direct numerical out-of-equilibrium
simulations~\cite{gia00} or by investigation of
soliton-soliton scattering processes~\cite{kho01}.
There has been, on the other hand, considerably
less progress for quantum systems. Huber, in
one of the firsts works on the subject~\cite{hub68},
evaluated the thermal conductivity $\kappa(T)$ for the
Heisenberg-chain with an equation-of-motion approximation
and found a {\em finite} $\kappa(T)$, a result which is,
by now, known to be wrong. It has been shown
recently~\cite{zot97}, that the energy-current operator
commutes with the hamiltonian for the
spin-1/2 Heisenberg chain. The thermal conductivity
is consequently {\em infinite} for this model.
The intriguing question, ``under which circumstances
does an interacting quantum-system show an
infinite thermal conductivity'', is to-date 
completely open, the analogous question for 
classical systems being intensively studied~\cite{bon00}.

A second motivation to study $\kappa(T)$ for quantum
spin-systems comes from experiment. An anomalous large
magnetic contribution to $\kappa$ has been 
observed~\cite{sol00} for the hole-doped spin-ladder system 
Sr$_{14-x}$Ca$_x$Cu$_{24}$O$_{41}$. For
Ca$_{9}$La$_{5}$Cu$_{24}$O$_{41}$, which has no
holes in the ladders, an even larger thermal
conductivity has been measured~\cite{hes01} raising
the possibility of ballistic magnetic transport limited
only by residual spin-phonon and impurity scattering.
There is, however, up-to-date no microscopic calculation
for the thermal conductivity of spin-ladders.

In this Letter we present  a finite size analysis of the 
thermal conductance in Heisenberg $J_1-J_2$ chains and
ladders suggesting ballistic thermal transport in these
two families of spin models. Following this result we propose a 
possible scenario that would account a number of 
anomalies found in the thermal conductivity in spin systems.   
We apply these ideas to the case of
$\rm Ca_9La_5Cu_{24}O_{41}$, obtaining 
good agreement with recent experimental measurements.

{\bf Models} -
We consider quasi one-dimensional systems for which the
hamiltonian takes the form $H=\sum_{{\rm x}=1}^L H_{\rm x}$,
where $L$ is the number of units along the chain.
We will consider two models, the isotropic
Heisenberg chain with dimerized nearest and homogeneous 
next-nearest neighbor exchange couplings,
\[
H_{x}^{(ch)} \ =\  J
\Big\{ (1+\delta(-1)^{\rm x})\,
{\bf S}_{\rm x} \cdot {\bf S}_{{\rm x}+1} 
+ \alpha\, {\bf S}_{\rm x} \cdot {\bf S}_{{\rm x}+2} \Big\}~,
\]
and the two-leg Heisenberg-ladder:
\[
H_{\rm x}^{(lad)} =
J_\parallel ({\bf S}_{1,\rm x}\cdot{\bf S}_{1,\rm x+1} 
+ {\bf S}_{2,\rm x}\cdot{\bf S}_{2,\rm x+1})
+ J_\perp {\bf S}_{1,\rm x}\cdot{\bf S}_{2,\rm x}~.
\]
$H^{(ch)}$ has been proposed to model the magnetic properties of the
spin-Peierls compound CuGeO$_{3}$. 
The  non-frustrated dimerized Heisenberg chain ($\alpha=0$)
models the magnetic behavior~\cite{gar97} of (VO)$_{2}$P$_{2}$O$_{7}$. 
Ca$_{9}$La$_{5}$Cu$_{24}$O$_{41}$ contains doped spin-chains and
undoped spin-ladders~\cite{hes01}, described by $H^{(lad)}$.

  
{\bf Method} -
Our goal is to make connection between the 
microscopic transport-properties of the low dimensional models 
presented above and the experimental measurements of 
thermal conductivity mentioned in the introduction.

The thermal conductivity is defined as the response 
of the energy-current density 
${\bf j}_{\bf x}^E$ to a thermal gradient
$\quad{\bf j}_{\bf x}^E = -\kappa\,\nabla T\quad$
and it has units of $[\kappa] = {\rm W\over K\, m^{d-2}}$,
where $d$ is the dimension ($d=3$ for experiments).
The specific form of the energy-current associated 
with a given hamiltonian is determined by the
continuity equation for the energy density,
$\dot H_{\rm x}+\nabla\cdot j_{\rm x}^E=0$, which leads
via $\nabla\cdot j_{\rm x}^E = \left(
j_{{\rm x}+1}^E-j_{\rm x}^E \right)/c$
to \cite{hei02}
\[
[H_{\rm x},H_{{\rm x}+1}]+ [H_{\rm x},H_{{\rm x}+2}]+
[H_{{\rm x}-1},H_{{\rm x}+1}] \ =\ 
{i\hbar\over c} \,j^{E}_{\rm x}~,
\]
where $c$ is the lattice constant along the chain-direction.

In the absence of applied magnetic fields spin-inversion
symmetry holds and the Kubo formula for 
$\kappa \ =\ \lim_{\omega\to0} \kappa(\omega)$
reduces to~\cite{hub68}
\begin{equation}  
\kappa \ =\ 
\lim_{\omega\to0} {\beta^2\over {\rm L}}
     \int_0^\infty dt\, {\rm e}^{t(i\omega-s)}
\langle J^E(t)J^E\rangle~,
\label{Kubo}
\end{equation}
where $J^E\ =\ \sum_{\bf x}\,{\bf j}_{\bf x}^E$ is the total 
energy-current, $\beta \ =1/(k_B T)$ is the inverse 
temperature and $s\to0$ the usual convergence factor.

We will examine here the possibility
of infinite intrinsic thermal conductivity. For such a system
heat transport is ballistic and energy is transported
without dissipation.
As an example we consider the XXZ chain
($H^{(ch)}$ with $\alpha=0=\delta$ and a spin-anisotropy).
The total thermal current commutes
with the hamiltonian in this case~\cite{zot97}
and the current-current correlation function 
is therefore time-independent.
Eq.\ (\ref{Kubo}) reduces then to the static expectation
value
$ \kappa(\omega=0)=\beta^2 \langle (J^E)^2 \rangle/(Ls)$,
a quantity which can be evaluated by Bethe-Ansatz~\cite{note_1}.
 
In general we have $[J^E,H] \neq 0$, but the thermal transport
will be ballistic if 
\begin{equation} 
\kappa^{(th)} \ = \  
\frac{\beta^2}{ZL}\sum_{m,n; E_n=E_m} 
e^{-\beta E_{m}} \left| \langle m|J^{E}|n \rangle\right|^{2}
\label{kappa_th}
\end{equation}
is finite in the thermodynamic limit $L\to\infty$.
The thermal conductivity
$\kappa \ =\ \kappa^{(th)}/s$ then diverges for $s\to0$;
as it does for the XXZ chain.
 
In reality, spin-hamiltonians like $H^{(ch)}$ 
and $H^{(lad)}$ are coupled to an external environment, e.g.\
to phonons or impurities.
Here we consider the case where this coupling is small.
This coupling will then result in a finite {\em external}
lifetime $\tau=1/s$ for the eigenstates of the spin-model
(which is assumed to be energy independent in Eq.\ (\ref{Kubo})).
Using the relation $\lambda = v_s\tau$ in between
the {\em external} mean-free path $\lambda$, the
spin-wave velocity $v_s$ and $\tau$, we propose
\begin{equation}  
\kappa \ =\ \kappa^{(th)}\, \tau
       \ =\ \kappa^{(th)}\,{\lambda\over v_s}
\label{Ballistic}
\end{equation}
to hold for the thermal conductivity $\kappa$.

Let us discuss the relation of Eq.\ (\ref{Ballistic})
to the usual phenomenological formula~\cite{Ashcroft}
(here in one dimension)
\begin{equation}  
\kappa^{(ph)}\ =\ c_V\, v_s\,\lambda~,
\label{Boltzmann}
\end{equation}
where $c_V$ is the specific heat. When applied
in order to analyze experimental data,
the microscopic equation (\ref{Ballistic}) will in general
yield different values for the magnetic mean-free path
$\lambda$. On the other hand, we might expect 
Eq.\ (\ref{Boltzmann}) to hold at low temperature
for the Heisenberg-chain in the gapless phase,
when the Luttinger-liquid quasiparticles are well
defined and a Boltzmann approach is justified.
Consequently we expect
$\kappa^{(th)}\,\lambda/ v_s \equiv c_V v_s\lambda$
in this case, i.e.\ we expect
$\kappa^{(th)}/ c_V \equiv v_s^2$ in the limit $T\to0$.
All three quantities in this equation 
($\kappa^{(th)}$, $c_V$ and $v_s$) can be computed
by Bethe-Ansatz for the XXZ-chain.
One finds that this equation is exact~\cite{note_1}
in the limit $T\to0$.

\begin{figure}[t]
\noindent
\\
\centerline{
\epsfig{file=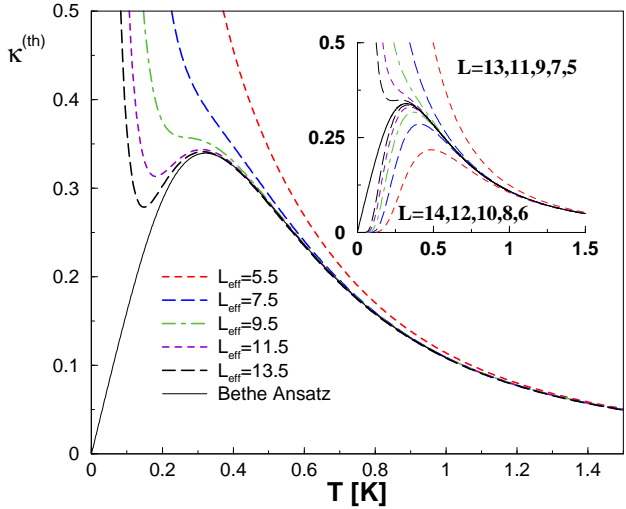,width=0.45\textwidth} 
}
\caption{\label{average}
$\kappa^{(th)}$, as defined by Eq.\ (\ref{kappa_th}),
for the Heisenberg-chain as a function of temperature.
Inset: Data for $L=5\dots14$, in comparison with
the Bethe-Ansatz result~\protect\cite{note_1}
(full line).
Main panel: Results using Eq.\ (\protect\ref{Average}),
with $L_{eff}=(L_0+L_1)/2$. We note, that the position of the
maximum in $\kappa^{(th)}$ can be determined confidently
with the averaging procedure.
}
\end{figure}

{\bf Numerics} -
The computation of  $\kappa^{(th)}(T)$ demands 
the  whole spectrum of the hamiltonian, restricting
the maximum lattice size and a careful finite-size
analysis is required.
In general one finds, e.g. for the XXZ-model
for which the exact Bethe-Ansatz result is 
known~\cite{note_1},
that the $\kappa^{(th)}$ in chains with   
odd(even) number of sites chains is a upper(lower) 
bound of the exact result. The value of  $\kappa^{(th)}(T)$  
decreases (increases) for chains with odd (even) number of sites 
when the lattice is enlarged. This observation has led
us to consider the average
\begin{equation} 
\kappa^{(th)}(L_{ \rm eff})\ =\ 
\frac{L_0\kappa^{(th)}(L_0)+L_1\kappa^{(th)}(L_1)}{L_0+L_1}
\label{Average}
\end{equation}
of finite-size data, where $L_0=2i$ and 
$L_1=2i+1$ and $L_{\rm eff}=(L_0+L_1)/2$.
We find that $\kappa^{(th)}(L_{ \rm eff})$
converges to the thermodynamic limit 
somewhat faster than taking the limit  
for even (or odd) number of sites only,
see Fig.\ (\ref{average}). This average technique is
reminiscent of the boundary-condition integration 
technique~\cite{gros92} for exact diagonalization studies. 

\begin{figure}[t]
\centerline{
\epsfig{file=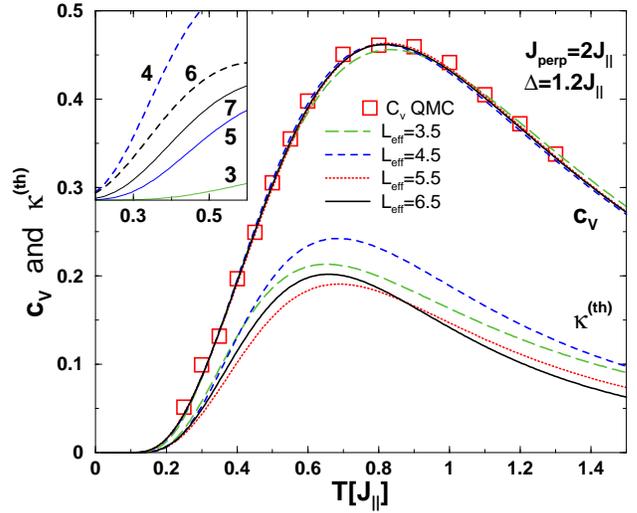,width=0.45\textwidth} 
           }
\vspace{4pt}
\caption{\label{lad_th}
  Temperature dependence of $\kappa^{(th)}$ and 
  the dimensionless specific heat for ladders with
  twisted boundary conditions and
  $J_{\perp}=2J_{\parallel} $.
  Both magnitudes are computed using the average procedure 
  (\protect\ref{Average}) described in the text. 
  The specific heat for a $2\times100$-ladder
  computed with QMC is plotted for comparison (squares).   
  Inset $\kappa^{(th)}$ for L=3,5,7 (from bottom up) and
  $L=4,6$ (from top down) in the low-temperature region.
}
\end{figure}

The situation is similar when boundary conditions are changed 
from periodic to  anti-periodic and for the ladders when the number 
of rungs changes from odd to even. We also find that (\ref{Average}) 
produces excellent results when applied to the specific 
heat, as illustrated in Fig.\ (\ref{lad_th}), where we
compare the exact diagonalization data for ladders
with Quantum-Monte-Carlo results~\cite{Alv00}.


{\bf Results} -
After these technical remarks we discuss now the 
exact diagonalization results for the temperature
dependence of $\kappa^{(th)}$.
The general features in all the models studied are: (i)
A finite value of $\kappa^{(th)}$ at any finite temperature that 
does not vanish when the extrapolation to infinite 
lattice size is taken. (ii) A single maximum $\kappa^{(th)}_{\rm max}$  
at an intermediate temperature $T_{max}(\kappa)$ smaller than the maximum
in the specific heat  $T_{max}(C_{v})$. (iii) 
The high temperature regime follows in general the law 
$C(L)/T^{2}$, as expected from Eq.\ (\ref{kappa_th}). 

The results for $H^{(lad)}$ are presented for
$J_\perp=2J_\parallel$ in Fig.\ (\ref{lad_th}).
We notice that the results for the specific heat are
already converged nicely, albeit the small
effective chain length used. We believe that the
results for $\kappa^{(th)}$ shown in Fig.\ (\ref{lad_th})
to be accurate to about $10\%$, enough to allow a
detailed analysis of the experimental data which we
will perform further below. In the inset of
Fig.\ (\ref{lad_th}) we present a blowup for
the data for $\kappa^{(th)}$ at low temperatures
for $L=3,\dots,7$. We note the systematic increase
of $\kappa^{(th)}$ with increasing $L$ for the
odd values $L=3,5,7$, indicating a finite
value in the thermodynamic limit.
The even values $L=4,6$ seem
to constitute upper bounds to $\kappa^{(th)}$.

\begin{figure}[t]
\noindent
\\
\centerline{
\epsfig{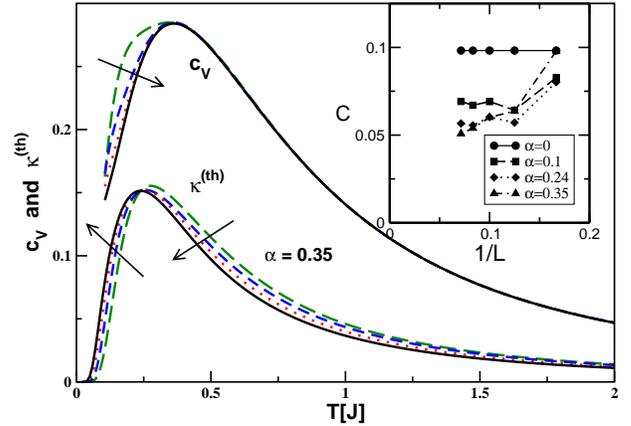} 
}
\caption{\label{frustrated}
$\kappa^{(th)}$ and dimensionless specific heat for the 
frustrated Heisenberg chain with $\alpha=0.35$
and different lattice sizes L=8,10,12,14, the arrows
indicate increasing $L$.
Inset : Finite size analysis (L=6...14)
of the high temperature residue $C(L)$ 
defined as  $\kappa^{(th)}= C(L)/T^{2}$ at  $T \gg J$
for different values of $\alpha$. 
}
\end{figure}

In Fig.\ (\ref{frustrated}) we present
the results for $H^{(ch)}$,
$\alpha=0.35$ and $\delta=0$. We have also
studied the dimerized phase with
$\delta>0$ and found similar behaviors.
The maximum value of  $\kappa^{(th)}$ is nearly
size-independent for L=8-14. The raise of
$\kappa^{(th)}$ with increasing
system size $L$ (as indicated by the arrows in
Fig.\ (\ref{frustrated})) in the low-temperature
regime, is consistent with
the notion of a finite $\kappa^{(th)}$
in the thermodynamic limit.

The way finite size chains approach the $L \to \infty$
limit is almost identical for frustrated chains 
and the exactly solvable $\alpha=0$ case, as
exemplified by the size-dependence for the prefactor of the
leading $1/T^2$-term presented in the
inset of Fig.\ (\ref{frustrated}),
showing the reliability of the finite size analysis. 
We can conclude that the frustration produces a 
substantial drop in the {\em extrapolated } finite value of 
the prefactor.
                               

{\bf Analysis} -  
Comparison with experimental results 
for the thermal conductivity $\kappa^{(exp)}$ 
can be made using the dimensional analysis:
\begin{equation} 
\kappa^{(exp)}\ =\ {k_B J\over \hbar c} 
         \left({N_c c^3\over a b c}\right)
         \left({\lambda\over c}\right)
         \left({J c\over \hbar v_s }\right)
	 \tilde \kappa^{(th)}
\label{1D3D}
\end{equation}  
where the quantities in the brackets are dimensionless.
$\tilde \kappa^{(th)}$ is the dimensionless thermal conductance
(\ref{kappa_th}) which we will evaluate by exact
diagonalization. 
$a$, $b$ and $c$ are the lattice constants
(the chains run along the c direction)
and $N_{c}$ is the number of chains per unit cell.
Note that
$\left({\lambda\over c}\right)
         \left({J c\over \hbar v_s }\right) =
         \left({J \tau\over \hbar }\right)
$
and (\ref{1D3D}) can be used to extract the lifetime
$\tau$ (in units of the coupling-constant $J$)
directly by comparison with the experimental
$\kappa^{(exp)}$.

\begin{figure}[t]
\centerline{
\epsfig{file=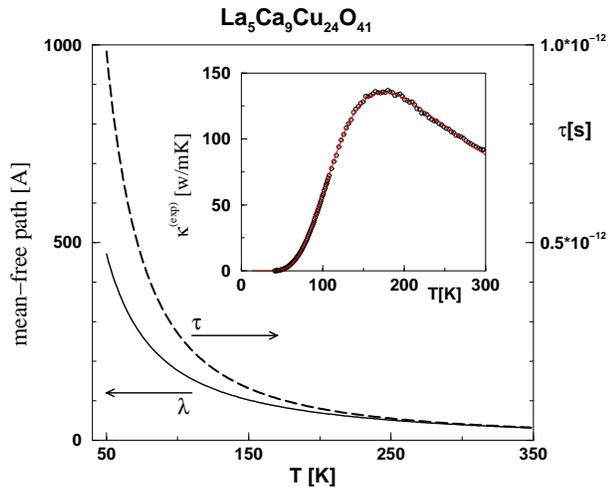,width=0.45\textwidth} 
           }
\vspace{4pt}
\caption{\label{La_exp}
$\tau(T)$ (right-scale) and $\lambda(T)$ (left-scale)
for Ca$_{9}$La$_{5}$Cu$_{24}$O$_{41}$ extracted using 
Eq.\ (\ref{1D3D}). The experimental data
by Hess {\it et al.}~\protect\cite{hes01} 
is shown in the inset (filled circles) together with an
analytic fit (solid line) discussed in the text.
}
\end{figure}

We now analyze the experimental data~\cite{hes01}
for Ca$_{9}$La$_{5}$Cu$_{24}$O$_{41}$. A
quantitative good description of the magnetic
excitations in La$_{9}$La$_{5}$Cu$_{24}$O$_{41}$ can 
be obtained by $H^{(lad)}$ with the inclusion
of an additional ring-exchange term~\cite{mat00}
(a 15\% correction) and 
$J_\perp\simeq J_\parallel$~\cite{mat00,win01}.
Here we disregard the possible ring-exchange term and
use $J\equiv J_\perp = J_\parallel=832.6\,{\rm K}$, which we
extracted by fitting the magnetic contribution~\cite{hes01}
$\kappa^{(exp)}(T)$ by
$444.2 / ( 1.0 + 0.093*\exp(419.6/T) + (T/156.4)^2 )$
(a very good fit at all temperatures, see inset of
Fig.\ (\ref{La_exp})).
The gap $\Delta$ for the isotropic Heisenberg ladder
is~\cite{kne01} $\Delta=0.504*J_\perp$, which leads for
Ca$_{9}$La$_{5}$Cu$_{24}$O$_{41}$ to
$J = 419.6/0.504\, {\rm K} =832.6\,{\rm K}$. 

Using the appropriate lattice constants and
$N_c=14$ (rungs per unit cell~\cite{McC88}) we used
Eq.\ (\ref{Average}) and (\ref{1D3D}) for 
$2\times6$ and $2\times7$ ladders in order to
extract the spin-environment relaxation time $\tau$, 
see Fig.\ (\ref{La_exp}). 
We find for the lifetime $\tau$ (in units of the coupling-constant $J$):
$\frac{\tau(T)\hbar}{J} \simeq 
7*\left(\frac{725}{T}+\left(\frac{561}{T}\right)^2
\right)$. The assumption of a weak spin-environment
coupling entering Eq.\ (\ref{Ballistic}) is therefore
justified in the experimental relevant temperature
regime. The lifetime is, to give an example,
132 times larger than the coupling constant $J$
at $T=150\,{\rm K}$, leading to an external 
broading of the energy levels of only
$1/132=0.76\%$, in units of $J$.

To estimate the effective mean-free path we use
$\lambda(T)=\bar v_s(T)\tau(T)$, were we have
used for $\bar v_{s}$ the thermal expectation 
value of the magnon-velocity within the
independent-triplet model:
\[
\bar v_{s} = {1\over Z} \int dk\, \varepsilon'(k)\, 
n(\varepsilon(k)) = {1\over Z} 
\int^{\Delta_{2}}_{\Delta} d\varepsilon\, n(\varepsilon)
\]
where $Z=\int dk\, n(\varepsilon(k))$ 
is the partition function and
$n(\varepsilon(k)) = 3/(\exp(\beta\varepsilon(k)+3)$.
We approximated
the one-magnon dispersion $\varepsilon(k)$ by
$2\,\varepsilon^2(k)=(\Delta^2+\Delta_{2}^2)
+ (\Delta_{2}^2-\Delta^2)\cos(k)$. For
$J_\perp=J_\parallel$ we have~\cite{kne01} that
$\Delta_{2}\approx3.8\Delta$.
In  the low-temperature limit
$\bar v_{s}(T)\approx\sqrt{T}$ holds and the mean-free
path $\lambda(T)$ is therefore less divergent at
low-$T$ than the relaxation time $\tau(T)$ see
Fig.\ (\ref{La_exp}). At very low temperature we 
expect impurity-scattering to become relevant and
$\lambda(T)$ to plateau-off.


{\bf Discussion} -  
Our results for the mean-free length $\lambda(T)$
for Ca$_{9}$La$_{5}$Cu$_{24}$O$_{41}$ 
obtained by the microscopic formula
(\ref{Ballistic}) are substantially smaller
than the one obtained using (\ref{Boltzmann}).
At $T=100\,{\rm K}$ Hess {\it et al}~\cite{hes01} estimated
a large $\lambda\approx 3000\AA$, 
in contrast to our result $\lambda(100)\approx 176\AA$. 
This is, in a certain sense,
surprising, since the effective $\lambda$ obtained
from (\ref{Boltzmann}) might be thought to contain
additional spin-spin scattering. Physically, the
reason for our smaller mean-free path stems from the
fact that $\kappa^{(th)}(T)$ peaks at substantially
smaller temperatures than $c_V(T)$, see
Fig.\ (\ref{frustrated}). We note, that all excitations
contribute to $c_V$ on an equal footing, but that
magnetic excitations near the bottom of the one-magnon
band seem to contribute dominantly to the thermal
conductivity, leading to a substantial reduction
of the temperature where $\kappa^{(th)}(T)$ is maximal,
with respect to $c_V(T)$.


{\bf Conclusions} -  
We have presented a microscopic approach of
thermal transport in quasi one-dimensional
spin models. For several models we find indications for
a diverging thermal conductivity. A finite
thermal conductivity is obtained when
couplings to external degrees of freedom are taken
into account with the relaxation-time approximation.
We have analyzed recent experiments for spin-ladder
compounds and found only weak
spin-environment coupling for the ladder compounds, which
decreases fast with decreasing temperature.
Our estimates for the
mean-free length turned out to be substantially smaller
than previous estimates using simple phenomenological 
formulas. These results highlights the importance of
microscopic theories for transport in quasi one-dimensional
quantum-spin systems.


{\bf Acknowledgments} -
This work was supported by the DFG. 
We thank T.~Lorenz, A.~Kl\"umper, B.~B\"uchner, W.~Brenig and G.S.~Uhrig
for discussions and for sending us their data. 
J.V.A. also thanks A.~Muramatsu and F.~Mila 
for stimulating discussions.




\end{document}